\documentclass{article}

\usepackage{PRIMEarxiv}

\usepackage[utf8]{inputenc} 
\usepackage[T1]{fontenc}    
\usepackage{hyperref}       
\usepackage{url}            
\usepackage{booktabs}       
\usepackage{amsfonts}       
\usepackage{nicefrac}       
\usepackage{microtype}      
\usepackage{lipsum}
\usepackage{graphicx}
\usepackage[numbers]{natbib}  
\usepackage{marvosym} 

\usepackage{amsmath,amssymb,amsfonts}
\usepackage{algorithmic}
\usepackage{graphicx}
\usepackage{textcomp}
\usepackage{xcolor}
\def\BibTeX{{\rm B\kern-.05em{\sc i\kern-.025em b}\kern-.08em
    T\kern-.1667em\lower.7ex\hbox{E}\kern-.125emX}}

\usepackage{url}
\usepackage{booktabs}
\usepackage{multirow,array}
\usepackage{rotating}

\newcommand{\shorttitle}{Hierarchical Exponential-Gaussian Mixtures for Watch-Time Distribution Prediction}

\hypersetup{
pdftitle={Hierarchical Exponential-Gaussian Mixtures for Watch-Time Distribution Prediction},
pdfsubject={
        cs.IR,          
        cs.LG,          
        cs.CY,          
        cs.MM,          
        cs.HC           
    },
pdfauthor={Sofia Gulevskaia, Mikhail Trapeznikov, Aleksandr Poslavsky, Alexander D'yakonov},
pdfkeywords={watch-time prediction, short-video recommendation, mixture density, variance collapse, distributional modeling, recommender systems, production deployment},
}

\begin{document}

\title{Hierarchical Exponential--Gaussian Mixtures for Watch-Time Distribution Prediction}

\author{
  Sofia Gulevskaia \hspace{0.5mm} \href{https://orcid.org/0009-0003-2266-3219} {\includegraphics[scale=0.06]{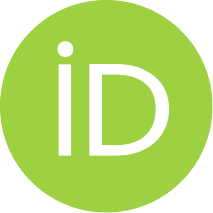}}\\
  AI VK \&
  Lomonosov MSU \\
  Moscow, Russia\\
  \texttt{s.gulevskaia@vkteam.ru} \\
  \And
  Mikhail Trapeznikov \hspace{0.5mm} \href{https://orcid.org/0009-0000-2939-0581} {\includegraphics[scale=0.06]{orcid.pdf}} \\
  AI VK \&
  Lomonosov MSU \\
  Moscow, Russia\\
  \texttt{m.trapeznikov@vk.team} \\
  \And
  Aleksandr Poslavsky \hspace{0.5mm} \href{https://orcid.org/0009-0009-5271-6696}{\includegraphics[scale=0.06]{orcid.pdf} }   \\
  AI VK\\
  Moscow, Russia\\
  \texttt{dr.slink@vk.com} \\
  \And
  Alexander D'yakonov \hspace{0.5mm} \href{https://orcid.org/0000-0001-7934-6538}{\includegraphics[scale=0.06]{orcid.pdf} \hspace{0.5mm}} \Letter \\
  AI VK\\
  Moscow, Russia\\
  \texttt{djakonov@mail.ru} \\
}

\maketitle

\begin{abstract}

Accurate watch-time (WT) prediction is an important requirement for
short-video recommendations.
Yet WT distributions are
near-zero-inflated, long-tailed and multimodal. The recent Exponential--Gaussian Mixture Network (EGMN) models the full conditional WT distribution rather than a single point estimate and achieves state-of-the-art performance.
Our large-scale reproduction study reveals
that EGMN is vulnerable to variance collapse, component redundancy, and inactive components.
We propose a Hierarchical Exponential--Gaussian Mixture (\textsc{HEGM}) model
that addresses these failure modes
through a hierarchical skip--watch decomposition, structured
initialization, removal of the forced Gaussian shift and entropy regularizer,
and optional KL-based variance regularization for over-parameterized mixtures.
Across large-scale industrial  and public datasets,
under a fixed-backbone
head-replacement protocol,
\textsc{HEGM} improves ranking accuracy and threshold-event prediction
over reproducible lightweight baselines,
while maintaining competitive point-estimation accuracy and substantially improving mixture stability and interpretability.
A 1.5-month production A/B test confirms statistically significant engagement lifts.
Our code and models are publicly released at \url{https://github.com/rw404/HEGM}.
\footnote{Accepted for publication at The IEEE International Conference on Data Mining (ICDM) 2026. This is a preprint version; the final version will appear in the conference proceedings.}
\end{abstract}


\section{Introduction}

\subsection{Motivation and Problem Significance}
Watch-time (WT) prediction is a production-critical component of modern short-video recommendation systems (e.g., TikTok, YouTube Shorts, Instagram Reels, and others), where full-screen, auto-playing feeds dominate content consumption~\cite{liu2026relative}. Within these ecosystems, conventional click-through signals become less informative because videos begin playing automatically after exposure, making click-through rate a misleading proxy for user satisfaction~\cite{covington2016deep, zhan2022deconfounding}.
Unlike sparse explicit feedback (e.g., likes, comments, shares), WT -- the duration $y \in \mathbb{R}_{+}$ that a user spends viewing a recommended video -- provides a dense, continuous measure of engagement.

\subsection{Key Challenges -- Why Point Prediction Is Insufficient}
WT, though observed as a scalar, exhibits a statistical structure that standard point-estimation objectives cannot capture.
Empirically, WT distributions in short-video platforms have several challenging properties~\cite{zhan2022deconfounding,sun2024cread,zhao2025multi}:
\textbf{near-zero inflation and right skewness} (many exposures result in very short WTs because users quickly scroll past content that does not capture immediate interest);
\textbf{long-tailed behavior} (rare high-engagement events contribute disproportionately to the upper tail);
\textbf{multimodality} (quick skips, partial views, completions, and replays correspond to distinct behavioral regimes).
In addition, WT is subject to various data biases -- such as duration bias: longer videos mechanically allow larger observed WTs and may therefore be over-favored by models trained directly on unadjusted labels~\cite{zhan2022deconfounding}.
Moreover, ranking consistency is more important than point calibration: the induced relative ordering of candidates matters more than absolute accuracy.
If two videos have true WTs $y_i$ and $y_j$, $y_i < y_j$, the model should produce scores satisfying $\hat{y}_i < \hat{y}_j$~\cite{zhan2022deconfounding, sun2024cread}.

\subsection{Distributional Watch-Time Modeling}
The limitations of point regression and discretization-based approaches motivate a shift toward conditional distribution estimation.
Rather than mapping a user--video--context instance $\mathbf{x}$ to a single scalar prediction $\hat{y}$, a distributional model estimates the full density $p_{\theta}(y \mid \mathbf{x})$ over the continuous WT variable.
This perspective is valuable for industrial recommendation systems because the same predictive distribution can support multiple ranking signals, including expected WT, threshold-crossing probabilities, completion probabilities, and uncertainty-aware decisions.

\subsection{EGMN}
A recent representative method in this direction is the Exponential--Gaussian Mixture Network (EGMN)~\cite{zhao2025multi}.
EGMN models WT using a mixture of an exponential component and several Gaussian components, reflecting the empirical structure of short-video consumption: a large mass of quick skips near zero and multiple engaged-watch regimes corresponding to partial views, completions, and replays.
The framework is \textbf{backbone-agnostic}, \textbf{integrates easily into multi-task production architectures, and provides various distributional statistics},
making it an attractive candidate for industrial deployment.


\subsection{Proposed Solution}
Despite its expressive potential, our reproduction study shows that the original EGMN is fragile at industrial scale.
In particular, maximum-likelihood training of unconstrained neural mixtures can lead to degenerate or poorly utilized components.
These failures also weaken  the ranking, watch-completion prediction, and interpretability properties required in production recommender systems.
We propose \textsc{HEGM} (\textbf{H}ierarchical \textbf{E}xponential--\textbf{G}aussian \textbf{M}ixture), a distributional WT prediction framework.
We preserve the main advantages of EGMN, while improving stability through hierarchical skip--watch decomposition, structured initialization, KL-based variance regularization and removing the forced Gaussian shift and the entropy regularizer. Our \textbf{main contributions} are as follows:

\begin{itemize}

\item \textbf{Empirical Diagnosis of Mixture Pathologies.} We provide the first detailed empirical study of failure modes in EGMN, demonstrating how failures in mixture training can cause the model to underperform simpler regressions.

\item \textbf{HEGM Proposal.} We introduce several principled modifications that substantially improve EGMN reliability and interpretability without sacrificing predictive power.

\item \textbf{Empirical Gains under a Fixed-Backbone Protocol.}
Across public benchmarks (KuaiRec, VK-LSVD) and a 282M-interaction industrial dataset, \textsc{HEGM}
outperforms reproducible lightweight head baselines in ranking quality and
threshold-event prediction.

\item \textbf{Production Deployment and Online Validation.}
A live A/B test on a major short-video platform confirms statistically
significant engagement lifts (e.g., $+9.26\%$ session depth), with
acceptable computational overhead.

\end{itemize}
We release code, preprocessing scripts, dataset splits, configurations, and checkpoints for reproducing all KuaiRec and VK-LSVD
experiments;
the industrial dataset and
online A/B-test data remain proprietary.

\section{Problem Formulation and Evaluation Metrics}
\label{sec:problem_formulation}


\subsection{Short-Video Watch-Time Prediction}
\label{sec:problem_wtp}

In a short-video recommendation setting, each logged impression is represented as a tuple
$(u_i, v_i, c_i, d_{v_i}, y_i)$, where $u_i$ is the user, $v_i$ is the video, $c_i$ is the exposure context, $d_{v_i} \in \mathbb{R}_{+}$ is the video duration, and $y_i \in \mathbb{R}_{+}$ is the observed WT. The model receives a feature vector
$\mathbf{x}_i = \phi(u_i, v_i, c_i, d_{v_i})$ and predicts user engagement for the corresponding impression.

In full-screen interfaces, videos play automatically upon impression, so $y_i$ is always defined.
Depending on logging policies, $y_i$ can either be strictly bounded by duration ($y_i \in [0, d_{v_i}]$) or exceed it ($y_i > d_{v_i}$) when repeated playbacks or loop counts are recorded.
In our work, all models are trained on the globally scaled target
$ y_i = y_i^{\mathrm{raw}} / s, $
where $s>0$ is a dataset-level scaling constant, such as a high percentile of WT. Unlike per-user, per-item, or per-duration normalization, this global scaling preserves absolute ranking information.

\subsection{Point and Distributional Prediction}
\label{sec:point_distribution}
Standard value-regression methods (VR) learn a scalar prediction
$\widehat{y}=f_{\theta}(\mathbf{x})$ and optimize a point-estimation loss such as \textbf{MAE}:
\begin{equation}
    \label{eq:MAE}
    \mathrm{MAE}=\frac{1}{n} \sum |y_i-\widehat{y}_i|,
    \end{equation}
or \textbf{MSE} ($\frac{1}{n} \sum (y_i-\widehat{y}_i)^2$).
Such objectives are efficient, but they reduce the conditional WT distribution to a single statistic and therefore discard information about skewness, multimodality, and uncertainty.
In contrast, distributional prediction estimates a conditional density $p_{\theta}(y \mid \mathbf{x})$ or its cumulative distribution function
$F_{\theta}(t \mid \mathbf{x})=\Pr(Y \leq t \mid \mathbf{x})$. 
From this distribution we obtain
\textbf{Expected WT (Point Estimate):}
$\widehat{y}_{\mathrm{mean}}(\mathbf{x}) = \mathbb{E}[Y \mid \mathbf{x}]$,
\textbf{Absolute Threshold Probability:}
$\Pr\nolimits_{\theta}(Y^{\mathrm{raw}} > \tau \mid \mathbf{x}) = 1 - F_{\theta}(\tau / s \mid \mathbf{x})$,
\textbf{Percentage Completion Probability:}
$\Pr\nolimits_{\theta}(Y^{\mathrm{raw}} > \rho d_v \mid \mathbf{x})$,
\textbf{Predictive Uncertainty:}
$\operatorname{Var}(Y \mid \mathbf{x}) = \mathbb{E}[Y^2 \mid \mathbf{x}] - \left(\mathbb{E}[Y \mid \mathbf{x}]\right)^2$.

\subsection{Evaluation Metrics}
\label{sec:problem_evaluation}

\subsubsection{Ranking Quality}
Recommender systems depend primarily on ordering consistency. We use \textbf{XAUC}~\cite{zhan2022deconfounding}, which estimates the probability that a predicted score $\widehat{y}$ correctly orders an interaction pair $(i, j)$ with $y_i \neq y_j$):
\begin{equation}
\operatorname{XAUC} = \frac{1}{|\mathcal{P}|} \sum_{(i,j)\in \mathcal{P}} \mathbf{1}\left[ (\widehat{y}_i-\widehat{y}_j)(y_i-y_j) > 0 \right],
\end{equation}
where $\mathcal{P}$ is a set of comparable pairs.

\subsubsection{Point accuracy}
We report \textbf{MAE} and \textbf{MSE} on the original WT scale, using
$\widehat y_i=\mathbb E[Y\mid\mathbf{x}_i]$ for distributional models.

\subsubsection{Distributional Quality} 
Negative log-likelihood (\textbf{NLL})
\begin{equation}
\label{eq:NNL}
\mathrm{NLL} = -\frac{1}{n} \sum \log p_{\theta}(y_i \mid \mathbf{x}_i)
\end{equation}
is the canonical scoring rule for continuous density estimation. However, we do not rely on NLL for model selection or as a primary metric, due to the variance collapse pathology discussed in Section~\ref{sec:prelim_egmn_weaknesses}.
This behavior is studied in this work.

\subsubsection{Threshold-event prediction}
Many production objectives are expressed through WT events, such as deep watch or percentage completion. For thresholds of the form $Y>\tau$ or $Y>\rho d_v$, we evaluate the corresponding binary predictions using \textbf{ROC AUC}~\cite{fawcett2006introduction}.

\section{Related Work}
\label{sec:related_work}

WT prediction for short-video recommendation has evolved from scalar point regression toward richer formulations that model the full conditional distribution of user engagement. 
We review the following relevant lines of work:


\subsection{Standard Regression and Classification}
It is natural to use dwell time for filtering or reweighting noisy click signals \cite{yi2014beyond, xie2023reweighting}. VR directly predicts WT by optimizing MSE or MAE.
While simple and computationally efficient, scalar regression compresses the heterogeneous, skewed, and multimodal conditional distribution $p(y \mid \mathbf{x})$ into a single statistic.
MSE corresponds to Gaussian maximum likelihood under a homoscedastic Gaussian observation model, an assumption clearly violated by real WT distributions.
Log-transforming the target can mitigate heavy tails, but retransformation to the original scale may introduce bias unless explicitly corrected~\cite{yu2026transun}.
WLR~\cite{covington2016deep} treats WT as a sample weight in a classification problem, but requires explicit positive and negative click labels, which are unavailable or unreliable in full-screen auto-playing scenarios.

\subsection{Ordinal and Discretization-Based Methods}
To capture ranking consistency, some works reformulate continuous WT prediction as classification tasks. TPM~\cite{lin2023tree} and PTPM~\cite{chen2025personalized} decompose prediction into ordered binary decisions. CREAD~\cite{sun2024cread} 
further improves this direction through Error-Adaptive Discretization (EAD) and restoration. Despite their effectiveness, discretization methods depend on bucket or tree design and lose part of the fine-grained continuous structure of WT.

\subsection{Debiasing and Causal Approaches}
WT is confounded by item popularity, noisy viewing behavior and duration bias.
To address this, D2Q~\cite{zhan2022deconfounding} predicts duration-dependent WT quantiles to isolate genuine user preference. DML~\cite{zhang2023leveraging} estimates the probability of exceeding WT quantiles conditional on membership in ``homogeneous'' groups -- for example, videos of roughly similar length.
Other approaches model the underlying components of WT explicitly: D2Co~\cite{zhao2023uncovering} uses a duration-wise Gaussian mixture model to separate meaningful interest from noise, while CWM~\cite{zhao2024counteracting} conceptualizes observed video WT as a truncation of an unobserved counterfactual watch-time (CWT) and introduces a correction function grounded in counterfactual analysis.
Similarly, RAD~\cite{liu2026relative} compares user behavior against contextual reference distributions.
These approaches are complementary to our work, which focuses on the stability and expressiveness of the predictive distribution head.


\subsection{Quantile and Uncertainty Modeling}
CQE~\cite{lin2024conditional} predicts multiple conditional quantiles via a pinball loss. While flexible, it requires selecting a fixed set of quantile levels and does not provide a continuous density. EXUM~\cite{wu2025explicit} incorporates an adversarial confidence head for explicit uncertainty control, whereas SWaT~\cite{yang2025swat} proposes a behavior-driven statistical approach that models continuation probabilities over progress-bar buckets. ProWTP~\cite{cui2025calibrating} offers WT calibration through prototype learning and optimal transport.

\subsection{Generative Modeling}
GR~\cite{ma2026generative} represents continuous values as sequences of discrete tokens in a positional numeral system and trains an autoregressive model. RQ-Reg~\cite{cui2026sequential} extends this via sequential prediction of coarse-to-fine residual codes. Although highly expressive, generative approaches 
introduce additional architectural and inference complexity compared with lightweight parametric heads.


\subsection{Exponential--Gaussian Mixtures}
The closest precursor to our work is EGMN~\cite{zhao2025multi}, which models WT as a mixture of one exponential component and several Gaussian components.
Section~\ref{sec:prelim_egmn} provides a detailed description.
We build HEGM upon EGMN while redesigning its parameterization and optimization
for stable mixture learning.

\section{Preliminaries: EGMN}
\label{sec:prelim_egmn}

We briefly review EGMN~\cite{zhao2025multi}, the state-of-the-art distributional WT model, which serves as the foundation for our work. EGMN treats WT as a conditional random variable rather than a scalar target to capture complex behavioral regimes, such as quick skips (mass near zero), partial views (multimodality), completions and replays (long tails).

\subsection{Exponential--Gaussian Mixture Distribution}
For the user-video-context feature vector $\mathbf{x} $ and observed WT $y$, let the conditional density $p(y \mid \mathbf{x})$ be a mixture of one exponential component and $K$ Gaussian components:
\begin{equation}
\begin{aligned}
    p_{\theta}(y \mid \mathbf{x})
    =
    w_{0}(\mathbf{x}) f_{\text{exp}}(y \mid \lambda(\mathbf{x})) \\
    +
    \sum_{k=1}^{K}
    w_{k}(\mathbf{x}) f_{\text{gauss}}(y \mid \mu_{k}(\mathbf{x}), \sigma_{k}^{2}(\mathbf{x})),
\end{aligned}
    \label{eq:egmn_density}
\end{equation}
where $w_{k}(\mathbf{x}) \ge 0$ and $\sum_{k=0}^{K}w_{k}(\mathbf{x})=1$,
\begin{align}
    f_{\text{exp}}(y \mid \lambda) &= \lambda e^{-\lambda y}, \quad y \geq 0, \label{eq:exp}\\
    f_{\text{gauss}}(y \mid \mu, \sigma^2) &= \frac{1}{\sqrt{2\pi\sigma^2}} \exp\left( -\frac{(y-\mu)^2}{2\sigma^2} \right). \label{eq:gauss}
\end{align}
The exponential component models coarse-grained skewness and quick skips near zero, while the Gaussians capture fine-grained engaged-watch patterns. The corresponding expected WT  $\widehat{y}_{\theta}(\mathbf{x})$ is available in closed form:
\begin{equation} 
    \mathbb{E}_{p_{\theta}}[Y \mid \mathbf{x}]
    =
    w_{0}(\mathbf{x})\lambda(\mathbf{x})^{-1}
    +
    \sum_{k=1}^{K}
    w_{k}(\mathbf{x})\mu_{k}(\mathbf{x}).
    \label{eq:egmn_expectation}
\end{equation}

\subsection{Network Architecture}
An underlying network $g_{\text{backbone}}$ first maps the features $\mathbf{x}$ to a latent representation $\mathbf{h} = g_{\text{backbone}}(\mathbf{x}) \in \mathbb{R}^{d}$.
EGMN operates as a backbone-agnostic prediction framework, with separate output heads that parameterize the mixture components:
\begin{equation}
\begin{aligned}
\lambda(\mathbf{x}) &= \operatorname{softplus}(\mathbf W_\lambda\mathbf h+\mathbf b_\lambda),\\
\sigma_k(\mathbf{x}) &= \operatorname{softplus}(\mathbf W_{\sigma_k}\mathbf h+\mathbf b_{\sigma_k}), \  k = 1,\ldots,K,\\
\mathbf w(\mathbf{x}) &= \operatorname{softmax}(\mathbf W_w\mathbf h+\mathbf b_w).
\end{aligned}
\end{equation}
To separate Gaussian components from the near-zero skip region, EGMN shifts each Gaussian mean by the exponential mean: for $k=1,\ldots,K$
\begin{equation}
    \mu_{k}(\mathbf{x})
    =
    \lambda(\mathbf{x})^{-1} + \operatorname{softplus}(\mathbf{W}_{\mu_k} \mathbf{h} + \mathbf{b}_{\mu_k}). 
    \label{eq:egmn_shifted_mu}
\end{equation}

\subsection{Training Objective}
EGMN is optimized via a composite loss function on a training set $\mathcal{D} = \{(\mathbf{x}_i,y_i)\}_{i=1}^{n}$:
\begin{equation}
    \mathcal{L}_{\mathrm{EGMN}}(\theta)
    =
    \mathcal{L}_{\mathrm{NLL}}(\theta)
    +
    \lambda_{\mathrm{ent}} \mathcal{L}_{\mathrm{ent}}(\theta)
    +
    \lambda_{\mathrm{reg}} \mathcal{L}_{\mathrm{reg}}(\theta),
    \label{eq:egmn_total_loss}
\end{equation}
where $\lambda_{\mathrm{ent}}, \lambda_{\mathrm{reg}} \ge 0$ are regularizing hyperparameters.
Minimizing the NLL loss $\mathcal{L}_{\mathrm{NLL}}(\theta)$ defined in~\eqref{eq:NNL} is equivalent to maximum likelihood estimation (MLE) and
aims to maximize the density assigned to the observed targets.
The entropy regularization term discourages weight degeneration by maximizing the
entropy of the mixture weights:
    \begin{equation}
    \label{eq:ent}
        \mathcal{L}_{\mathrm{ent}}(\theta) = \frac{1}{n} \sum_{i=1}^{n} \sum_{k=0}^{K} w_{k}(\mathbf{x}_i) \log w_{k}(\mathbf{x}_i).
    \end{equation}
The regression term $\mathcal{L}_{\mathrm{reg}}$ is the standard MAE loss~\eqref{eq:MAE}.

\section{Limitations and Design Rationale}
\label{sec:prelim_egmn_weaknesses}

\begin{figure}[tbp]
    \centering
    \includegraphics[width=0.75\linewidth]{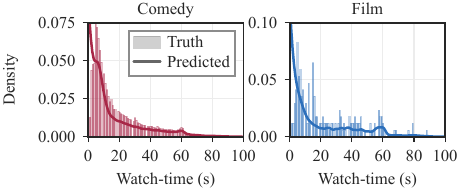}
    \caption{Watch-time distributions: histograms from real data and predicted densities (via \textsc{HEGM}). Histograms are shown for different
         video categories (comedy and film).}
    \label{fig:empirical_distributions}
\end{figure}

To confirm the motivation behind distributional modeling, we analyze WT patterns on our industrial data. As noted in prior papers, the conditional distribution of WT varies significantly across user cohorts and video categories. Fig.~\ref{fig:empirical_distributions} illustrates this heterogeneity and the densities predicted by our \textsc{HEGM}.
Because individual-level predicted
densities are noisy and difficult to inspect, we report the average density
within each group (e.g., per user cohort or per video category), which yields
smoother, more interpretable visualizations.

Our large-scale reproduction and industrial evaluation expose several failure modes inherent to EGMN. 

\subsection{Gaussian Variance Collapse}
\begin{figure*}[tbp]
    \centering
    \includegraphics[width=0.97\linewidth]{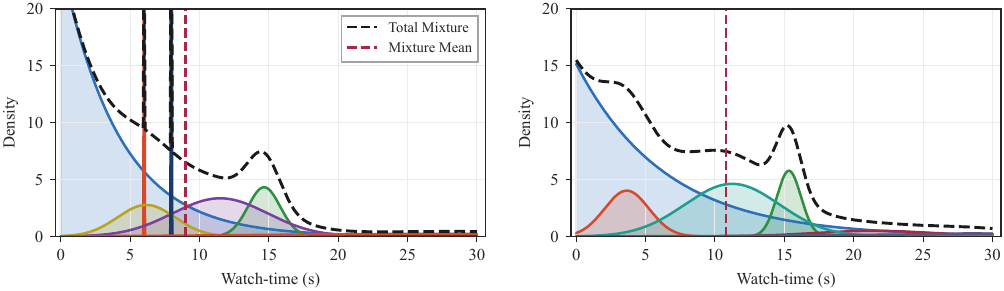}
    \caption{Predicted densities from EGMN with collapsed components (left) and from \textsc{HEGM} after stabilization (right).}
    \label{fig:PIC3_delta}
\end{figure*}

Maximum-likelihood training of Gaussian mixtures can produce degenerate components with $\sigma_k(\mathbf{x}) \rightarrow 0$, causing likelihood spikes around individual samples rather than robust density estimates~\cite{bishop2006pattern, mclachlan2000finite}. 
In EGMN, this pathology leads to example memorization, unstable uncertainty estimates and poor component interpretability. Fig.~\ref{fig:PIC3_delta} illustrates the resulting delta-like ``needle'' densities (here and in some subsequent density plots, WT is shown on the original, unnormalized scale).
Our proposed modifications (Section~\ref{sec:method}) eliminate these artifacts (right panel).

\subsection{Component Redundancy \& Inactivity}
\begin{figure}[tbp]
    \centering
    \includegraphics[width=0.75\linewidth]{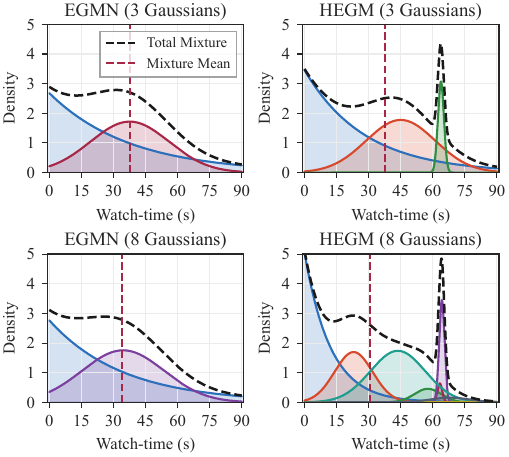}
    \caption{Predicted mixtures from EGMN (left column) and \textsc{HEGM} (right column). EGMN underutilizes multiple components, while \textsc{HEGM} preserves a more interpretable multimodal structure.
       The rightmost peak aligns with video duration, capturing complete views.}
    \label{fig:PIC2_mixture}
\end{figure}

EGMN often fails to use its Gaussian components effectively: several Gaussians may merge ($\mu_i \approx \mu_j, \sigma_i \approx \sigma_j$) into a single component centered near the exponential mean $1 / \lambda$. As a result, the predicted distributions remain nearly identical regardless of the number of components $K$ (see Fig.~\ref{fig:PIC2_mixture}, left), and the effective modeling capacity does not increase with $K$. 
Moreover, the distribution sometimes degenerates to a pure exponential form ($w_0(\mathbf{x}) \approx 1$), losing multimodality entirely. Attempts to mitigate these issues inadvertently produce delta-like spikes as a side effect.
Sometimes, despite the entropy regularization loss, the mixture weights for all but one or two Gaussians consistently collapse to near zero ($w_k \approx 0$).
We sought a solution that 
maintains a clearer multimodal structure with well-separated components
and without collapses; see Fig.~\ref{fig:PIC2_mixture} (right).

\subsection{Poor Production Performance \& Initialization Sensitivity}
With a basic EGMN implementation, we achieved only modest performance on our production data. The model underperforms simple scalar MSE regression (XAUC $0.6585$ vs $0.7070$, see Section~\ref{sec:exp_main_results}), and the results are sensitive to initialization.

These failure modes directly motivate our proposed \textsc{HEGM} framework.

\section{Proposed Method: \textsc{HEGM}}
\label{sec:method}

The goal of HEGM is not to introduce a generic new mixture-learning
principle, but to make Exponential--Gaussian WT modeling robust under the
behavioral structure of short-video feeds.

We outline the key differences between \textsc{HEGM} and EGMN.

\subsection{Hierarchical Skip--Watch Decomposition}
\begin{figure}[tbp]
    \centering
    \includegraphics[width=0.75\linewidth]{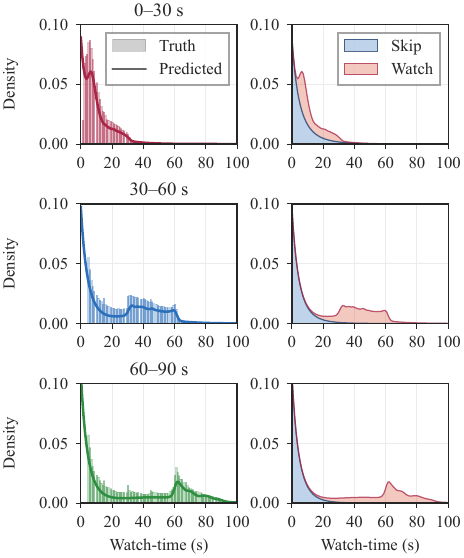}
    \caption{Watch-time distribution decomposition. Histograms for different duration buckets (left column): 0\,--\,30\,s, 30\,--\,60\,s, 60\,--\,90\,s, and separation into exponential component and Gaussian mixture (right column).}
    \label{fig:PIC4_decomposition}
\end{figure}

Instead of
EGMN's flat softmax mixture,
\textsc{HEGM} separates quick-skip behavior from engaged watching through an explicit two-stage behavioral hierarchy:
a sigmoid gate estimates the skip probability $p_{\text{skip}}(\mathbf{x}) = \sigma(\mathbf{W}_{\text{skip}} \mathbf{h} + b_{\text{skip}}) \in (0,1)$ based on the hidden representation $\mathbf{h} = g_{\text{backbone}}(\mathbf{x}) \in \mathbb{R}^d$.
The full predictive density is
\begin{equation}
\begin{aligned}
    p_{\theta}(y \mid \mathbf{x})
    =
    p_{\text{skip}} (\mathbf{x}) \cdot f_{\text{exp}}(y \mid \lambda(\mathbf{x})) \\
    +
    (1 - p_{\text{skip}} (\mathbf{x})) \cdot \sum_{k=1}^{K}
    w_{k}(\mathbf{x}) f_{\text{gauss}}(y \mid \mu_{k}(\mathbf{x}), \sigma_{k}^{2}(\mathbf{x})),
\end{aligned}
    \label{eq:egmn_density2}
\end{equation}
the weights $w_{k}(\mathbf{x}) > 0$ satisfy $\sum_{k=1}^{K} w_{k}(\mathbf{x}) = 1$ and are obtained via softmax. From an analytical perspective, this decomposition provides interpretable separation between \textit{lack of interest} (quick skips) and \textit{active engagement} (actual viewing). Fig.~\ref{fig:PIC4_decomposition} demonstrates this for videos grouped by duration buckets. The clear separation of the exponential and Gaussian components enhances both model interpretability and downstream analytics.


\subsection{Global Scaling and Structured Initialization}
Recall that in  our work before modeling we use a global scaling: $y=y^{\mathrm{raw}}/s$, $s > 0$. To reduce component merging and stabilize the first epochs of training, \textbf{Gaussian means} are initialized
at evenly spaced locations in the interval:
\begin{equation}
        \mu_{k}^{(0)} = \frac{k}{K+1} \ \  \text{or} \ \  \mu_k \sim \text{Uniform}(0.1, 0.9),
    \end{equation}
$k=1,\dots,K$ (in this work we use deterministic uniform initialization),  and \textbf{Gaussian standard deviations} are initialized as
$\sigma_{k}^{(0)} =  2 / (3 K)$
(to a scale preserving early separation while avoiding spikes), \textbf{the exponential component} is initialized using a short-watch prior ($0.05 \approx 9\text{\,s}$ on industrial scale):
$\lambda^{(0)} = 1 / 0.05$.

\subsection{Removal of Forced Gaussian Shift}
Unlike EGMN, \textsc{HEGM} does not shift Gaussian means by the exponential mean \eqref{eq:egmn_shifted_mu}. We instead parameterize them directly, allowing engaged-watch components to adapt freely to the empirical distribution:
\begin{equation}
    \mu_{k}(\mathbf{x})
    =
    \operatorname{softplus}(\mathbf{W}_{\mu_k} \mathbf{h} + \mathbf{b}_{\mu_k}). 
\end{equation}
EGMN's forced shift couples skip and watch regimes: \(\lambda(x)\) models
near-zero mass, but \(1/\lambda(x)\) also shifts all Gaussian means. Thus,
skip-heavy gradients can indirectly move engaged-watch components and promote
redundant Gaussians near the shared offset. HEGM removes this coupling and
lets the two regimes adapt independently.

\subsection{Removal of Entropy Regularization}
We also remove entropy regularization over mixture weights~\eqref{eq:ent}, as we found empirically that it did not improve model performance.
Entropy regularization 
encourages mixture weights to remain active,
but it does not separate
component means or control variances. Consequently, active components may still
be redundant, and poorly placed Gaussians may form sharp local spikes. Hence,
weight entropy does not reliably prevent collapse or improve ranking.

\subsection{Optional KL-Based Variance Regularization for Collapse Control}
For large $K$, Gaussian components can still collapse ($\sigma_k(\mathbf{x}) \rightarrow 0$). We therefore introduce an optional variance prior that penalizes degenerate components without constraining their means.
The scaled video duration range is divided into equal-width buckets
$\mathcal{B}=\{B_1,\ldots,B_M\}$. For each bucket $b$, we compute a reference variance from the training data:
$\bar{\sigma}_{b}^{2} =  \operatorname{Var}\left( y_i : d_{v_i}\in B_b \right)$.
Let $b(i)$ denote the bucket containing video $v_i$. The KL penalty is
\begin{align}
    \mathcal{L}_{\mathrm{KL}}
    &=
    \frac{1}{nK} \sum_{i=1}^{n} \sum_{k=1}^{K} D_{\mathrm{KL}}\left( \mathcal{N}\left(\mu_{k}, \sigma_{k}^{2}(\mathbf{x}_i)\right) \,\middle\|\, \mathcal{N}\left(\mu_{k}, \bar{\sigma}_{b(i)}^{2}\right) \right) \nonumber\\
    &=
    \frac{1}{2nK} \sum_{i=1}^{n} \sum_{k=1}^{K} \left[ \frac{\sigma_{k,\theta}^{2}(\mathbf{x}_i)}{\bar{\sigma}_{b(i)}^{2}} - \log \frac{\sigma_{k,\theta}^{2}(\mathbf{x}_i)}{\bar{\sigma}_{b(i)}^{2}} - 1 \right].
    \label{eq:variance_kl}
\end{align}
The penalty anchors component variances to duration-conditioned empirical scales,
naturally accounting for the higher variance of longer clips while discouraging
both degenerate spikes and excessively diffuse components.
Critically, it is mean-agnostic: by penalizing only the variance, it preserves the
ability of different components to model distinct engagement modes.

\subsection{Training Objective}
The final \textsc{HEGM} objective is
\begin{equation}
    \mathcal{L}_{\mathrm{HEGM}} = 
    \mathcal{L}_{\mathrm{NLL}} + \lambda_{\mathrm{reg}}\mathcal{L}_{\mathrm{reg}} + \lambda_{\mathrm{KL}}\mathcal{L}_{\mathrm{KL}}.
    \label{eq:hegm_total_loss}
\end{equation}

\section{Experiments}
\label{sec:experiments}


\subsection{Experimental Setup}
\label{sec:exp_setup}

\subsubsection{Datasets and Preprocessing}
We use three short-video recommendation datasets (see Table~\ref{tab:datasets}):
\begin{itemize}
    \item \textbf{KuaiRec}~\cite{gao2022kuairec}: A dense, public benchmark, a de facto standard in WT prediction~\cite{zhao2025multi, cui2026sequential, ma2026generative, yang2025swat, sun2024cread}. \url{https://kuairec.com}
    \item \textbf{VK-LSVD}~\cite{poslavsky2026vk}: A large-scale public VK video dataset; we use the \texttt{up0.01\_ir0.01} subsample, which is similar to public benchmarks. \url{https://huggingface.co/datasets/deepvk/VK-LSVD}
    \item \textbf{Industrial (Internal)}: Our primary testbed for ranking
      quality, compiled from a production short-video platform serving
      millions of active users. The dataset is proprietary, but the public VK-LSVD dataset was collected from the same platform.
\end{itemize}

\subsubsection{Global Scaling} 
Constants for global scaling are reported in Table~\ref{tab:datasets}.
We scale the Industrial dataset following our production convention, while VK-LSVD and KuaiRec are scaled by their respective empirical 99th percentiles of video duration.
We do not apply per-user, per-item, or per-duration normalization.

\begin{table*}[tbp]
    \centering
    \caption{Dataset statistics and corresponding hyperparameter configurations. } 
    \label{tab:datasets}
    \begin{tabular}{lrrrrrcccc} 
        \toprule
        \textbf{Dataset}
        & \textbf{Users}
        & \textbf{Videos}
        & \textbf{Interactions}
        & \textbf{Norm. scale $s$}
        & \textbf{Train\,/\,Val\,/\,Test}
        & \textbf{Learning rate}
        & \textbf{Dropout} \\ 
        \midrule
        KuaiRec & $7{,}176$ & $10{,}728$ & $12{,}530{,}806$ & $60$\,s & 80\,/\,10\,/\,10  (\%) & $10^{-5}$ & 0.3 \\ 
        VK-LSVD & $100{,}000$ & $171{,}106$ & $38{,}404{,}921$ & $157$\,s & 5\,/\,1\,/\,1 (weeks) & $10^{-4}$  & 0.1 \\ 
        Industrial & $6{,}390{,}738$ & $2{,}484{,}528$ & $282{,}443{,}525$ & $180$\,s & 5\,/\,1\,/\,1 (days) & $10^{-4}$ & 0.1 \\ 
        \bottomrule
    \end{tabular}
\end{table*}
\subsubsection{Data splitting}
Unlike many works that employ random splitting in WT prediction, all datasets are split chronologically to avoid look-ahead bias and mirror real production scenarios; the exact split definitions are reported in
Table~\ref{tab:datasets}.

\subsubsection{Backbones and Features}
We took special care to ensure that our comparison was not disadvantaged by implementation or tuning differences.
To guarantee a fair evaluation, all heads share the same underlying backbones.
For public datasets, numerical and categorical features use
PLE~\cite{gorishniy2022embeddings} and embedding layers, respectively, followed
by a parallel DCNv2~\cite{wang2021dcn} stack with a residual decoder and
linear prediction heads.
The full implementation is provided in the accompanying code.

For VK-LSVD, count features are Laplace-smoothed. We additionally train
three iALS models on (i)~positive feedback, (ii)~raw WT, and (iii)~WT $\ge15$\,s.
From the resulting user and item embeddings, we construct dot-product and cosine-similarity features, as well as the $\ell_2$-norms of the user and item embeddings, ensuring no target leakage.


KuaiRec preprocessing follows the feature setup of~\cite{zhao2025multi} (user--video interaction features were not used).

For the Industrial dataset, the production backbone is a multi-task
architecture with wide logits, user embeddings, and task-specific towers
for clicks, likes, shares, WT, etc.

\subsubsection{Baselines}
Our goal is not to compare all possible WT predictors under their native architectures, but to evaluate replacement heads under a fixed industrial ranking backbone. We compare \textsc{HEGM} against reproducible, backbone-agnostic baselines that integrate into identical shared hidden states:
\begin{itemize}
    \item \textbf{MSE-VR:} Direct value regression optimized via MSE.
    \item \textbf{MAE-VR:} Direct value regression optimized via MAE.
    \item \textbf{CREAD:} The
strongest non-mixture baseline in the EGMN comparison~\cite{zhao2025multi}; we took the implementation from the public EGMN repository: \url{https://github.com/BestActionNow/EGMN/}.
    \item \textbf{EGMN:} The
closest distributional predecessor~\cite{zhao2025multi}. 
\end{itemize}


\subsubsection{Hyperparameters and Training Details}
All models are trained with Adam~\cite{kingma2014adam} for up to $30$ epochs, using a batch size of $4096$ and an embedding dimension of $8$ across all datasets. 
Dataset-specific hyperparameters are listed in
Table~\ref{tab:datasets}.
For mixture models, we tune the number of Gaussian components $K \in \{3,6,9,12\}$, 
the KL regularization weight
$\lambda_{\mathrm{KL}} \in \{0.05, 0.1, 0.5, 1.0\}$,  and the regression loss weight $\lambda_{\mathrm{reg}} \in \{0.5, 1.0, 2.0\}$. 
We use $M=18$ duration buckets for KL-based variance regularization.

\subsubsection{Model Selection}
All offline models and hyperparameter values are selected exclusively on the validation split. The primary selection metric is validation XAUC because the production ranker is optimized for ordering quality. 
All reported metrics are computed on the test set only after selecting the checkpoint.
No test-set feedback is used for hyperparameter selection.

Unless stated otherwise, results are averaged over three random seeds.

\begin{table*}[ht]
\centering
\caption{Experimental results on KuaiRec, VK-LSVD, and Industrial datasets.
         Best values for each metric and dataset are boldfaced.}
\scriptsize
\label{tab:main_results}
\begin{tabular}{@{}l@{\hskip 8pt} cccc @{\hskip 16pt} cccc @{\hskip 16pt} cccc@{}}
\toprule
& \multicolumn{4}{c}{\textbf{XAUC} $\uparrow$}
& \multicolumn{4}{c}{\textbf{MAE} (raw WT) $\downarrow$} 
& \multicolumn{4}{c}{\textbf{MSE} (raw WT) $\downarrow$}  \\
\cmidrule(lr){2-5} \cmidrule(lr){6-9} \cmidrule(lr){10-13}
\textbf{Model}
& $K=3$ & $K=6$ & $K=9$ & $K=12$
& $K=3$ & $K=6$ & $K=9$ & $K=12$
& $K=3$ & $K=6$ & $K=9$ & $K=12$ \\
\midrule
\multicolumn{13}{@{}l}{\textbf{KuaiRec}} \\ 
\midrule
MSE-VR   & \multicolumn{4}{c}{0.5467} & \multicolumn{4}{c}{4.65}  & \multicolumn{4}{c}{59.17} \\
CREAD      & \multicolumn{4}{c}{0.5616} & \multicolumn{4}{c}{4.54}  & \multicolumn{4}{c}{\textbf{58.84}} \\
EGMN       & 0.5504 & 0.5541 & 0.5584 & 0.5587 & 4.54 & 4.52 & \textbf{4.43} & \textbf{4.43} & 59.38 & 59.31 & 60.44 & 60.29 \\
\textbf{HEGM} & \textbf{0.5622} & 0.5598 & 0.5595 & 0.5567 & 4.49 & 4.51 & 4.50 & 4.51 & 59.29 & 59.34 & 59.29 & 59.40 \\
\midrule
\multicolumn{13}{@{}l}{\textbf{VK-LSVD}} \\ 
\midrule
MSE-VR   & \multicolumn{4}{c}{0.6148} & \multicolumn{4}{c}{15.89} & \multicolumn{4}{c}{484.15} \\
CREAD      & \multicolumn{4}{c}{0.6193} & \multicolumn{4}{c}{\textbf{15.56}} & \multicolumn{4}{c}{480.61} \\
EGMN       & 0.6173 & 0.6163 & 0.6151 & 0.6152 & 16.07 & 16.07 & 15.96 & 16.01 & 480.70 & 481.11 & 483.66 & 484.76 \\
\textbf{HEGM} & 0.6202 & \textbf{0.6206} & 0.6197 & 0.6202 & 15.70 & 15.74 & 15.77 & 15.77 & 480.05 & \textbf{478.92} & 479.87 & 479.68 \\
\midrule
\multicolumn{13}{@{}l}{\textbf{Industrial}} \\ 
\midrule
MSE-VR       & \multicolumn{4}{c}{0.7070} & \multicolumn{4}{c}{13.12} & \multicolumn{4}{c}{\textbf{569.99}} \\
MAE-VR       & \multicolumn{4}{c}{0.6557} & \multicolumn{4}{c}{14.34} & \multicolumn{4}{c}{764.08} \\
CREAD          & \multicolumn{4}{c}{0.7146} & \multicolumn{4}{c}{24.14} & \multicolumn{4}{c}{1380.23} \\
EGMN           & 0.6585 & 0.6501 & 0.6507 & 0.6558 & 13.51 & 13.92 & 13.75 & 14.04 & 730.74 & 726.88 & 744.05 & 731.01 \\
\textbf{HEGM}   & \textbf{0.7188} & 0.7178 & 0.7176 & 0.7162 & \textbf{12.97} & 13.07 & 13.07 & 13.04 & 578.61 & 583.75 & 583.68 & 580.63 \\
\textbf{HEGM+KL} & 0.7179 & 0.7168 & 0.7168 & 0.7168 & 13.08 & 13.07 & 13.05 & 13.08 & 585.08 & 584.92 & 581.98 & 584.54 \\
\bottomrule
\end{tabular}
\end{table*}

\subsection{Main Results and Convergence Behavior}
\label{sec:exp_main_results}

Table~\ref{tab:main_results} reports the main offline results across all datasets. 
For \textsc{HEGM} $\lambda_{\text{KL}} = 0$, for \textsc{HEGM+KL} $\lambda_{\text{KL}} = 0.1$.
MAE and MSE are reported for the raw target (before scaling).
Because the chronological data split is fixed across runs, variation due to data splitting is eliminated. Across three random seeds, the resulting standard deviations of XAUC
are below 
$0.0001$, so we omit them from Table~\ref{tab:main_results} for brevity. The key observations are as follows:

\begin{itemize}
    \item HEGM achieves superior XAUC -- our primary metric -- across all datasets compared to all baselines. Notably, CREAD achieves better performance than EGMN.
    \item For MAE, different models (CREAD, EGMN, HEGM) achieve the best results on different datasets. Our model performs best on the industrial dataset.
    \item For MSE, CREAD performs best on KuaiRec, whereas HEGM achieves the best result on VK-LSVD. On the Industrial dataset, MSE-VR remains best. We attribute this to the complex production backbone, the large data volume, and direct MSE optimization by MSE-VR. However, MSE regression is inferior in terms of XAUC ($0.7070 \rightarrow 0.7188$).
    \item CREAD achieves strong XAUC on the Industrial dataset ($0.7146$), but its MAE and MSE are substantially worse than those of MSE-VR and HEGM, because CREAD optimizes a classification-restoration objective over discretized watch-time buckets. HEGM is more suitable when the same head must support ranking, threshold probabilities, and downstream analytics based on the predictive distribution.
    \item HEGM also often outperforms EGMN in terms of NLL ($-2.38$ vs. $-1.84$ on the industrial dataset). However, NLL is excluded from Table~\ref{tab:main_results} because variance collapse renders it unstable and unreliable for interpretation.
    
\end{itemize}

\textsc{HEGM} improves ranking while remaining competitive in point-estimation metrics.
The best configuration on Industrial requires only a sparse allocation of $K=3$ Gaussian components, showing that excess components are unneeded when the structural framework is regularized effectively.

\begin{figure}[tbp]
    \centering
    \includegraphics[width=0.75\linewidth]{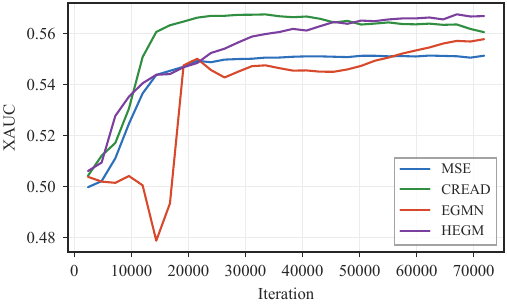}
    \caption{Validation curves of the compared methods on KuaiRec.}
    \label{fig:lcurve}
\end{figure}

Fig.~\ref{fig:lcurve} shows representative validation trajectories. CREAD may converge quickly, then overfit.
Mixture-density heads require more epochs to stabilize component locations, weights, variances, and the induced point expectation.
\textsc{HEGM} ($K=3$) continues improving in later epochs, overtaking CREAD
only after $\approx 20$ epochs, suggesting that short early-stopping
schedules may underestimate distributional architectures.

\subsection{Watch-Completion Analysis}
\label{sec:exp_threshold}

Table~\ref{tab:threshold_auc} reports ROC AUC for predicting whether the
observed WT exceeds a given threshold on the Industrial dataset. We evaluate
ROC AUC against binary targets derived from two families of thresholds:
absolute time thresholds ($> T$ seconds) and duration-normalized thresholds
($> P\%$ of video duration). For point baselines, we use predicted WT as the score; for distributional models, we use the corresponding exceedance probability computed directly from the predicted distribution.
No additional fine-tuning or head modification is performed; the same predictive distribution is reused directly for these auxiliary tasks.

Methods that model the full conditional distribution -- rather than a single point estimate -- demonstrate clear advantages on these metrics. This finding highlights the inherent benefit of distributional prediction for downstream threshold-based tasks, as a single predictive density supports multiple decision criteria without retraining. \textsc{HEGM} variants outperform all baselines, and the margins widen
substantially for deep absolute engagement tasks: at the $60$-second threshold,
\textsc{HEGM} achieves $0.8903$ ROC AUC compared to EGMN's $0.8356$.

Notably, \textsc{HEGM+KL} performs well in this evaluation,
more consistently outperforming the unregularized \textsc{HEGM} than in the
point-prediction and ranking metrics. We attribute this to the KL-based variance regularizer, which stabilizes
component variances and produces more reliable tail densities, directly
benefiting threshold-based decisions at extreme quantiles.

\begin{table}[tbp]
    \centering
    \caption{ROC AUC $\uparrow$ for watch-completion (Industrial dataset).}
    \label{tab:threshold_auc}
    \setlength{\tabcolsep}{2pt} 
    \begin{tabular}{lcccccc}
        \toprule
        \textbf{Threshold}
        & \textbf{MSE-VR}
        & \textbf{MAE-VR}
        & \textbf{CREAD}
        & \textbf{EGMN}
        & \textbf{\textsc{HEGM}}
        & \textbf{\textsc{HEGM}+KL} \\
        \midrule
        $10$~s & 0.7419 & 0.7139 & 0.7520 & 0.7502 & 0.7592 & \textbf{0.7593} \\
        $30$~s & 0.8050 & 0.7130 & 0.8088 & 0.8063 & 0.8341 & \textbf{0.8344} \\
        $60$~s & 0.8476 & 0.6862 & 0.8483 & 0.8356 & 0.8903 & \textbf{0.8907} \\
        $10\%$ & 0.8182 & 0.8218 & 0.8107 & 0.8190 & \textbf{0.8511} & 0.8477 \\
        $25\%$ & 0.6892 & 0.6951 & 0.6927 & 0.6954 & \textbf{0.7021} & 0.6998 \\
        $50\%$ & 0.6564 & 0.6572 & 0.6702 & 0.6741 & \textbf{0.6758} & 0.6747 \\
        $75\%$ & 0.6563 & 0.6592 & 0.6730 & 0.6868 & 0.6957 & \textbf{0.6969} \\
        $90\%$ & 0.6573 & 0.6615 & 0.6737 & 0.6907 & 0.7076 & \textbf{0.7103} \\
        \bottomrule
    \end{tabular}
\end{table}

\subsection{Ablation Studies}
\label{sec:exp_ablation}
We conduct ablations on the Industrial dataset  to assess the contribution of
the main design choices of \textsc{HEGM}.
Starting from the basic model with $K=3$, specific structural elements are systematically stripped or reverted.
Table~\ref{tab:ablation} reports metrics for the \textsc{HEGM} model variants.

\begin{table}[tbp]
    \centering
    \caption{Ablation study on the Industrial dataset ($K=3$).}
    \label{tab:ablation}
    \begin{tabular}{lccc}
        \toprule
        \textbf{Variant} & \textbf{XAUC} $\uparrow$ & \textbf{MAE} $\downarrow$ & \textbf{MSE} $\downarrow$ \\
        \midrule
        \textbf{\textsc{HEGM}} & \textbf{0.7188} & \textbf{12.97} & \textbf{578.61} \\
        \quad w/o structured initialization & 0.6841 & 13.54 & 644.15 \\
        \quad w/o sigmoid (flat mixture) & 0.7164 & 13.05 & 583.80 \\
        \quad with forced mean shift & 0.7171 & 13.08 & 585.42 \\
        \quad with entropy regularization & 0.7153 & 13.08 & 586.71 \\
        \quad with KL regularization, $\lambda_{\text{KL}} = 0.1$ & 0.7179 & 13.08 & 585.08 \\
        EGMN& 0.6585 & 13.51 & 730.74 \\
        \bottomrule
    \end{tabular}
\end{table}

Notably, structured initialization proves critical: removing it causes a sharp drop in XAUC from $0.7188$ to $ 0.6841$.
Restoring the flat mixture, forced Gaussian shift, or entropy regularization degrades all reported metrics, supporting our decision to remove them.
KL regularization slightly reduces ranking accuracy at $K=3$, but remains useful for better watch-completion predictions (Section~\ref{sec:exp_threshold}) and
prevents variance collapse in larger mixtures (Section~\ref{sec:exp_collapse}).

\begin{figure*}[tbp]
    \centering
    \includegraphics[width=0.97\linewidth]{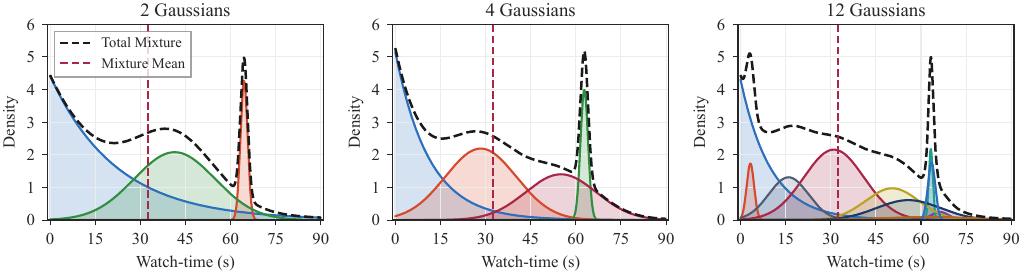}
    \caption{Predicted mixtures produced by \textsc{HEGM} for varying component counts $K$.}
    \label{fig:PIC6_mixtures}
\end{figure*}

\textit{Varying Component Count $K$.}
On the Industrial dataset, \textsc{HEGM} and \textsc{EGMN} achieve stable performance across different values of 
$K$ (Table~\ref{tab:main_results}).
Fig.~\ref{fig:PIC6_mixtures} shows predicted mixtures (from our HEGM approach) of WT distributions for different numbers of components. We show illustrative examples for $K = 2, 4, 12$, which were not used
for model selection and are chosen for visual clarity.

\subsection{Collapse Prevention}
\label{sec:exp_collapse}

We quantify Gaussian collapse by measuring the fraction of predicted components whose standard deviation satisfies $\sigma_k(\mathbf{x}) < \epsilon$. In Table~\ref{tab:collapse} the offline statistics demonstrate several complementary structural effects:
\begin{itemize}
    \item For small mixtures ($K=3,\,6$), \textsc{HEGM} eliminates variance collapse ($0\%$), whereas EGMN defaults to spikes ($5\%$–$10\%$). This component memorization is associated with poor generalization.
    \item  When the number of components becomes large, unregularized \textsc{HEGM} can still collapse (collapse rate $47.88\%$ at $K=12$ under $\epsilon=10^{-3}$).
    \item  Adding KL variance regularization $\mathcal{L}_{\mathrm{KL}}$, $\lambda_{\text{KL}} = 0.1$ eliminates collapse under the tested $\epsilon$ thresholds, suggesting that KL regularization is primarily needed for over-parameterized mixtures.
\end{itemize}

\begin{table}[tbp]
    \centering
    \caption{Percentage of collapsed Gaussians: $\sigma < \epsilon$ (Industrial dataset)}
    \label{tab:collapse}
    \begin{tabular}{llrrrr}
        \toprule
        \textbf{Method} & \textbf{$\boldsymbol{\epsilon}$} & \textbf{$K=3$} & \textbf{$K=6$} & \textbf{$K=9$} & \textbf{$K=12$} \\
        \midrule
        {EGMN} 
        & $10^{-4}$ & $5.10\%$ & $6.10\%$ & $7.55\%$ & $8.18\%$ \\
        & $10^{-3}$ & $8.95\%$ & $9.73\%$ & $12.24\%$ & $12.69\%$ \\
        \midrule
        {\textsc{HEGM}} 
        & $10^{-4}$ & 0.00\% & 0.00\% & $1.07\%$ & $22.33\%$ \\
        & $10^{-3}$ & 0.00\% & 0.00\% & $11.07\%$ & $47.88\%$ \\
        \midrule 
        \textsc{HEGM}+KL & $10^{-4}$ & 0.00\% & 0.00\% & 0.00\% & 0.00\% \\
        $\lambda_{\text{KL}} = 0.1$ & $10^{-3}$ & 0.00\% & 0.00\% & 0.00\% & 0.00\% \\
        \bottomrule
    \end{tabular}
  
\end{table}

\begin{figure}[tbp]
    \centering
  \includegraphics[width=0.75\linewidth]{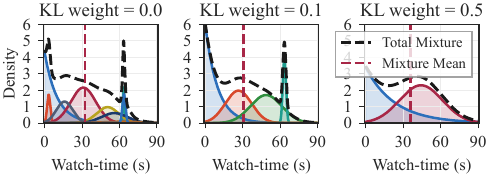}
    \caption{Predicted watch-time densities for identical user--video contexts: from collapse when $\lambda_{\text{KL}}=0$ to a smoother multimodal structure when $\lambda_{\text{KL}}>0$.}
    \label{fig:collapse-visualization}
\end{figure}

\begin{figure}[tbp]
    \centering
    \includegraphics[width=0.75\linewidth]{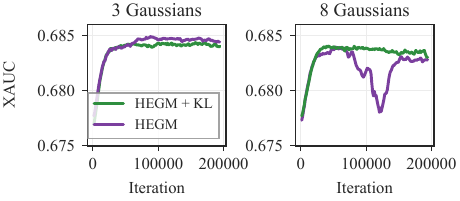}
    \caption{Training dynamics of \textsc{HEGM} with and without KL regularization. Left: $K=3$ components. Right: $K=8$ components.}
    \label{fig:PIC8_curves}
\end{figure}

Fig.~\ref{fig:collapse-visualization} provides a qualitative comparison:
increasing $\lambda_{\text{KL}}$ eliminates spikes and smooths the density.
Fig.~\ref{fig:PIC8_curves} further shows that KL regularization improves training stability when many components are used. For $K=3$ (left panel), both variants (HEGM and HEGM+KL) behave similarly; for larger mixtures ($K=8$, right panel), the unregularized model shows stronger XAUC fluctuations. We observed the same behavior for EGMN (see Fig.~\ref{fig:lcurve}).

\section{Online Deployment and A/B Testing}
\label{sec:exp_online}

We deployed \textsc{HEGM} as a drop-in replacement for the proprietary
multi-task WT prediction head in a production ranking system.
The retrieval pipeline and feature generation
were kept unchanged, isolating the effect of the proposed distributional
head.
The deployed configuration used $K=3$,
$\lambda_{\mathrm{reg}}=1.0$, and $\lambda_{\mathrm{KL}}=0$ (since collapse
was absent for $K=3$).

All experiments were conducted on an HPC cluster equipped with
$4 \times$ NVIDIA H100 GPUs (80\,GB each).
A single training run completes within
approximately $8$ hours; we did not track total GPU-hours across all
experiments.
The serving infrastructure for the online A/B test uses
$48 \times$ NVIDIA RTX 6000 Pro GPUs (96\,GB each).
All models are implemented in PyTorch~2.9 with CUDA~12.3 and deployed
in TensorRT format for production inference.

\subsection{Experimental Design}
The experiment was conducted as a user-level randomized A/B test over a period
of 1.5 months. A total of $5\%$ of live traffic was routed to the
\textsc{HEGM} treatment group and $5\%$ to the production baseline, with the
remaining traffic unaffected. This allocation corresponds to approximately
$11$ million requests per day in each group. To isolate the causal effect of the model change from temporal confounds, we
employed a forward--reverse testing protocol: a forward A/B test was run for
the first month, followed by a two-week reverse test in which the treatment
and control assignments were swapped.
The sign and magnitude of the effects were consistent across both phases,
which reduces the likelihood that the observed lifts are driven by temporal
seasonality or novelty effects. We distinguish two complementary criteria for interpreting the results:
\begin{itemize}
    \item \textbf{Statistical significance} was assessed via CUPED-adjusted $t$-tests applied to user-level aggregates. All reported metrics satisfy $p < 0.05$ after the CUPED variance reduction.
    \item \textbf{Practical business significance} was evaluated against operational thresholds that reflect mature platform scales and the minimum effect size considered practically meaningful by the product team; the full list is provided in Table~\ref{tab:online_ab}.
\end{itemize}

\subsection{Online A/B Test Results}
Table~\ref{tab:online_ab} reports the average relative changes of
\textsc{HEGM} against the production baseline in the forward A/B test.
All primary engagement metrics improved, with most exceeding their practical
significance thresholds.
Session depth increased substantially (by $+9.26\%$ from $28.52$ to
$31.16$ videos per session), dislikes decreased, and shares increased.
Total view time remained largely unchanged.
This pattern suggests that the gains are not driven by increasing total
consumption time, but by reducing immediate skips and increasing the number
of meaningful watch events per session.
The improvements in threshold-based watch metrics did not translate into a comparable lift in likes,
which is consistent with explicit feedback and consumption signals capturing complementary aspects
of user preference.

\begin{table}[tbp]
    \centering
    \caption{Forward-phase online A/B results for \textsc{HEGM} relative to the production WT head.
Bold: statistically significant ($p<0.05$) and above the predefined
business significance threshold.}
    \label{tab:online_ab}
    \begin{tabular}{lcc}
        \toprule
        \textbf{Primary Engagement Metric} & \textbf{Relative Change} & \textbf{Threshold}  \\
        \midrule
        Non-short watch ($Y \ge 3$\,s) & \textbf{+\,4.59\%} & $3\%$ \\
        Deep watch ($Y \ge 10$\,s) & \textbf{+\,5.75\%} & $3\%$ \\
        Skips ($Y < 3$\,s) & \textbf{--\,6.23\%} & $4\%$ \\
        Session depth (videos/session) & \textbf{+\,9.26\%} & $5\%$ \\
        \midrule
        \multicolumn{2}{l}{\textbf{Platform Guardrail Metric}} \\
        \midrule
        Total view time (TVT) & --\,0.03\% & $1\%$ \\
        Likes & +\,0.10\%  & $3\%$ \\
        Dislikes & \textbf{--\,8.34\%}  & $5\%$\\
        Shares & \textbf{+\,14.59\%} & $8\%$ \\
        \bottomrule
    \end{tabular}
\end{table}

\subsection{Computational Overhead}
Resource metrics evaluated over 1B production requests (Table~\ref{tab:latency}) confirm that
\textsc{HEGM}
adds 1.22\,ms to median latency; p99 rises from 7.9\,ms to 8.2\,ms,
still far below the 30\,ms serving constraint.

\begin{table}[tbp]
    \centering
    \caption{Production inference cost.}
    \label{tab:latency}
    \setlength{\tabcolsep}{5pt}
\begin{tabular}{lcccc}
\toprule
\textbf{Metric} & \textbf{Baseline} & \textbf{\textsc{HEGM}} & \textbf{Delta} & \textbf{Constraint} \\
\midrule
Latency (median) & 1.83\,ms & 3.05\,ms & +\,1.22\,ms & 30\,ms \\
99th percentile latency    & 7.9\,ms  & 8.2\,ms  & +\,0.3\,ms  & 30\,ms \\
Model parameters           & 1.3\,M   & 1.5\,M   & +\,0.2\,M   & --- \\
Batch size                 & 16{,}000 & 16{,}000 & 0         & --- \\
\bottomrule
\end{tabular}
\end{table}

\section{Conclusion}

This paper is motivated by a broader shift in WT modeling: recent methods increasingly move beyond scalar point prediction and fixed discretization toward distributional, uncertainty-aware, quantile-based, generative, and regime-based formulations. Within this landscape, EGMN represents a principled step toward distributional WT modeling.

Our contribution is not a generic mixture learner, but a targeted redesign of EGMN: we diagnose failure modes in its parameterization, fix them with minimal overhead, and validate the result through ablation, offline tests, and production deployment. Code and models are publicly released.


Under a fixed-backbone evaluation, HEGM achieved the best XAUC across all datasets,
improved threshold-event prediction, and eliminated variance collapse for small mixtures; KL variance regularization further stabilized over-parameterized mixtures.
A 1.5-month production A/B test produced significant engagement gains, including a 9.26\% increase in session depth, with modest inference overhead.

Several limitations remain.
First, \textsc{HEGM} improves predictive distribution modeling but does not
explicitly address causal biases such as duration or exposure bias.
Second, its Exponential--Gaussian family remains a parametric approximation;
alternative component distributions may better reflect the nonnegative and
potentially heavy-tailed support of WT.
Third, although we analyze likelihood behavior and component collapse,
a more comprehensive evaluation of probabilistic calibration and sharpness
is warranted.
Finally, our explanation of EGMN failure modes is primarily empirical;
formal conditions for collapse under shifted or entropy-regularized mixtures
remain an open problem.

Overall, our results suggest that the practical success of distributional WT models depends not only on expressive mixture families, but also on parameterizations that prevent component degeneracy and are aligned with user behavior.


\bibliographystyle{IEEEtran}
\bibliography{WT_links}

\end{document}